1# AWSOM-LP: An Effective Log Parsing Technique Using Pattern Recognition and Frequency Analysis

Issam Sedki, Abdelwahab Hamou-Lhadj, *Senior Member IEEE,* and Otmane Ait-Mohamed, *Member IEEE,*
PREPRINT
This work has been submitted to the IEEE for possible publication. Copyright may be transferred without notice, after which this version may no longer be accessible.

**Abstract**—Logs provide users with useful insights to help with a variety of development and operations tasks. The problem is that logs are often unstructured, making their analysis a complex task. This is mainly due to the lack of guidelines and best practices for logging, combined with a large number of logging libraries at the disposal of software developers. There exist studies that aim to parse automatically large logs. The main objective is to extract templates from samples of log data that are used to recognize future logs. In this paper, we propose AWSOM-LP, a powerful log parsing and abstraction tool, which is highly accurate, stable, and efficient. AWSOM-LP is built on the idea of applying pattern recognition and frequency analysis. First, log events are organized into patterns using a simple text processing method. Frequency analysis is then applied locally to instances of the same group to identify static and dynamic content of log events. When applied to 16 log datasets of the the LogPai project, AWSOM-LP achieves an average grouping accuracy of 93.5%, which outperforms the accuracy of five leading log parsing tools namely, Logram, Lenma, Drain, IPLoM and AEL. Additionally, AWSOM-LP can generate more than 80% of the final log templates from 10% to 50% of the entire log dataset and can parse up to a million log events in an average time of 5 minutes. AWSOM-LP is available online as an open source. It can be used by practitioners and researchers to parse effectively and efficiently large log files so as to support log analysis tasks.

## 1 INTRODUCTION

SOFTWARE logging is an important activity that is used by software developers to record critical events that would help them analyze a system at runtime. Logs are generated by logging statements inserted in the source code. An example of a logging statement is shown in Figure 1, which is a code snippet extracted from a Hadoop Java source file.

The generated log event (Figure 1) is composed mainly of two parts: the log header and log message. The log header typically contains the timestamp, the log level, and the logging function. The log message contains static tokens (i.e., usually text) such as "Received block", "size of", "from" in the example of Figure 1 and dynamic tokens, which represent variable values. In the example of Figure 1, we have three variable values, which represent the block id (blk_-5627280853087685), block size (67108864), and an IP address (/10.251.91.84).

Log files contain a wealth of information regarding the execution of a software system used to support various software engineering tasks including anomaly detection [1]

- I. Sedki is with the Department of ECE, Concordia University, Montreal, QC, Canada.
  E-mail:i_sedki@live.concordia.ca
- A. Hamou-Lhadj is with the Department of ECE, Concordia University, Montreal, QC, Canada.
  E-mail: wahab.hamou-lhadj@concordia.ca
- O. Ait-Mohamed is with the Department of ECE, Concordia University, Montreal, QC, Canada.
  E-mail: otmane.aitmohamed@concordia.ca

[2] [3], debugging and understanding system failure [4] [5] [6] [7] [8] , anomaly detection [1] [9] [10] [11] [12], testing [13], performance analysis [14] [15] [16], automatic analysis of logs during operation [17] [8] [14] [18] [19], failure prediction [8] [20], data leakage detection [12] and recently for AI for IT Operations (AIOps) [21] [17]. Logs, however, have been historically difficult to work with. First, typical log files can be significantly large (in the order of millions of events) [22] [23] [7] . Add to this, the practice of logging remains largely ad hoc with no known guidelines and best practice [24]. There is no standardized way for representing log data either [23]. To make things worse, logs are rarely structured [1] [20], making it difficult to extract meaningful information from large raw log files [25] [26].

In this paper, we focus on the problem of log parsing and abstraction, which consists of automatically converting unstructured raw log events into a structured format that would facilitate future analysis. Several log parsing and abstraction tools have been proposed (e.g., [27] [28] [29]). More precisely, log parsing techniques focus on parsing the log message. This is because log headers usually follow the same format within a log file. Parsing a log message is further reduced to the problem of automatically distinguishing the static text from the dynamic variables. The result of parsing the log event of Figure 1 consists of extracting the template shown in Figure 1, where the structure of the log message is clearly identified. A simple way to parse log events is to use regular expressions [30] [31]. The problem is that there may exist thousands of such templates in industrial log files

```
Logging statement        : LOG.info("Received Block"+ block + "of size" +
block_size + " from" + sender_ip)

Raw log line             : 081109 203519 147 INFO dfs.DataNode$PacketResponder:
Received block blk_−1608999687919862906 of size 91178 from /10.250.14.224

Log template : Received Block <*> of size <*>  from <*>
```

Fig. 1. A code snippet showing an example of a logging statement, the generated log event, and the expected log template

[16] [32]. In addition, as the system evolves, new log formats are introduced due to the use of various logging libraries, requiring constant updates of the regular expressions.

In this paper, we propose AWSOM-LP[1], a powerful automated log parsing approach and tool that leverages pattern recognition and frequency analysis. AWSOM-LP starts by identifying patterns of log events using similarity measurements and clusters them into groups. It then applies frequency analysis to instances of each group to differentiate between static and dynamic tokens of log messages. The idea is that tokens that are repeated more frequently are most likely static tokens than variable values. This is not the first time that frequency analysis is used in log parsing. Logram, a recent approach proposed by Dai et al. [27], also uses frequency analysis. However, Logram applies frequency analysis to the entire log file, which makes it hard to find clear demarcation between static and dynamic tokens. AWSOM-LP, on the other hand, applies frequency analysis to log events that belong the same pattern, which increases the likelihood of distinguishing between static and dynamic tokens. In addition, AWSOM-LP does not require building 3-gram tables as it is the case for Logram, which greatly simplifies the parsing process approach.

We evaluated AWSOM-LP using 16 log datasets from the LogPai benchmark[2] and compared it to five leading log parsing tools, mainly Drain [29], AEL [31], Lenma [28], IPLoM [33] and Logram [27]. Our results show that AWSOM-LP performs better than these tools in parsing 13 out of the 16 log datasets. In addition, our approach has an average accuracy of 93.5%, whereas the second best method, Logram, has an average accuracy of 82.5%. Additionally, AWSOM-LP can parse large files in a few minutes. When applied to 12 log files, it took AWSOM-LP less than 5 min to parse up to 1 million log events. For small files (100 thousand events and less), the average parsing time is less than 1 minute. AWSOM-LP is also stable. It requires between 10% to 50% of the log data to learn 80% of the templates.

AWSOM-LP is available online as an open source[3]. It can be used by practitioners and researchers to parse large log files to support the various log analysis techniques.

---

1. AWSOM-LP stands for **A**bdel**W**ahab Hamou-Lhadj, I**S**sam Sedki, and **O**t**M**ane Ait-Mohamed Log Parser.
2. https://github.com/logpai
3. https://github.com/SRT-Lab/awsom-lp

**Paper organization.** The paper is organized as follows. Section 2 introduces the background of log parsing and surveys prior work in that area. Section 3 presents, through a simple cherry-picked running example, a comprehensive description of AWSOM-LP approach. Section 4 shows the outcomes of assessing AWSOM-LP's accuracy, efficiency and ease of stabilization. Section 5 discusses the limitations and the threats to the validity of our findings. Finally, Section 6 concludes the paper.

## 2 RELATED WORK

Log analysis has received a great deal of attention from researchers and practitioners in recent years, due to the increasing need to understand complex systems at runtime. Log parsing is an essential prerequisite for log analysis tasks. Perhaps the most comprehensive survey of log parsing techniques is the one proposed by El-Masri et al. [24] in which the authors proposed a quality model for classifying log parsing techniques and examined 17 different log parsing techniques tools using this model. Existing tools can be categorized into groups based on the techniques they use, namely rule-based parsing tools, frequent tokens mining, natural language processing, and classification/clustering approaches. Another excellent study on surveying log parsing tools include the study of Zhu et al. [30]. We discuss the main approaches in what follows and conclude with a general comparison of AWSOM-LP with these techniques.

Jiang et al. [34] introduced AEL (Abstracting Execution Logs), which is a log parsing method that relies on textual similarity to group log events together. AEL starts by detecting trivial dynamic using hard-coded heuristics based on system knowledge (e.g., IP addresses, numbers, memory addresses). The resulting log events are then tokenized and assigned to bins based on the number of terms they contain. This grouping is used to compare log messages in each bin and abstracts them into templates. The problem with AEL is that it assumes that events that contain the same number of terms should be grouped together, resulting in many false positives.

Vaarandi et al. [35], [36] proposed SLCT (Simple Logfile Clustering Tool). The authors used clustering techniques to identify log templates. SLCT groups log events together based on their most common frequent terms. To this end, the approach relies on a density-based clustering algorithm to recognize the dynamic tokens, SLCT uses frequency analysis across all log lines in the log file. LogCluster [21] is an improved version of SLCT proposed by the same authors. LogCluster extracts all frequent terms from the log messages and arranges them into tuples. Then, it splits the log file into clusters that contain at least a certain number of log messages ensuring that all log events in the same cluster match the pattern constructed by the frequent words and the wildcards, which replace the dynamic variables.

Another clustering approach is the one proposed by Makanju et al., which is called IPLOM (Iterative Partitioning Log Mining) [33]. IPLOM employs a heuristic-based hierarchical clustering algorithm. Using this approach, the first step is to partition log messages. For this, the authors used heuristics considering the size of log events. The next step is to further divide each partition based on the highest number of similar terms. The resulting leaf nodes of the hierarchical partitioning as clusters and event types. Fu et al. proposed LKE (Log Key Extraction) [37], another clustering-based approach, using a distance-based clustering technique. Log events are grouped together using weighted edit distance, giving more weight to terms that appear at the beginning of log events. Then, LKE splits the clusters until each raw log level in the same cluster belongs to the same log key, and extracts the common parts of the raw log key from each cluster to generate event types. Tang et al. proposed LogSig [38], which considers the words present in a log event as signatures of event types. LogSig identifies log events using a set of predefined message signatures. First, it converts log messages into pairs of terms. Then, it forms log-message clusters using a local search strategy. LogSig selects the terms that appear frequently in each cluster and use them to build the event templates.

Hamooni et al. proposed LogMine [39], which uses MapReduce to abstract heterogeneous log messages generated from various systems. The LogMine algorithm consists of a hierarchical clustering module combined with pattern recognition. It uses regular expressions based on domain knowledge to detect a set of dynamic variables. Then, it replaces the real value of each field with its name. It then clusters similar log messages with the friends-of-friends clustering algorithm.

Natural Language Processing (NLP) techniques have also been used for log parsing. Logram, a recent approach proposed by Dai et al. [27], is an automated log parsing approach developed to address the growing size of logs, and the need for low-latency log analysis tools. It leverages n-gram dictionaries to achieve efficient log parsing. Logram stores the frequencies of n-grams in logs and relies on the n-gram dictionaries to distinguish between the static tokens and dynamic variables. Moreover, as the n-gram dictionaries can be constructed concurrently and aggregated efficiently, Logram can achieve high scalability when deployed in a multi-core environment without sacrificing parsing accuracy. In Logram, the identification of dynamic and static tokens depends on a threshold applied to the number of times the n-grams occur. An automated approach to estimates this threshold was proposed. Kobayashi et al. proposed NLP-LTG (Natural Language Processing–Log Template Generation) [40], which considers event template extraction from log messages as a problem of labeling sequential data in natural language. It uses Conditional Random Fields (CRF) [41] to classify terms in log messages as a static or dynamic. To construct the labeled data (the ground truth), it uses human knowledge and regular expressions.

Thaler et al. proposed NLM-FSE (Neural language Model-For Signature Extraction) [42], which trains a character-based neural network to classify static and dynamic parts of log messages. The approach constructs the training model through four layers. (1) The embedding layer transforms the categorical character input into a feature vector. (2) The Bidirectional-LSTM layer [43] allows each prediction to be conditioned on the complete past and future context of a sequence. (3) The dropout layer avoids over-fitting by concatenating the results of the bi-LSTM layer, and (4) the fully connected, feed-forward neural network layer predicts the event template using the Softmax activation function.

He et al. [29] proposed Drain, a tool that abstracts log messages into event types using a parse-tree. Drain algorithm consists of five steps. Drain starts by pre-processing raw log messages using regular expressions to identify trivial dynamic tokens, just like AWSOM-LP. Then, it builds a parse-tree using the number of tokens in log events. Drain assumes that tokens that appears in the beginning of a log message are most likely static tokens. It uses a similarity metric that compares leaf nodes to event types to identify log groups.

Spell (Streaming Parser for Event Logs using an LCS) is a log parser, which converts log messages into event types. Spell relies on the idea that log messages that are produced by the same logging statement can be assigned a type, which represents their longest common sequence. The LCS of the two messages is likely to be static fields.

The main difference between AWSOM-LP and existing approaches lies in the way AWSOM-LP is designed. AWSOM-LP leverages the idea that static and dynamic tokens of log events can be better identified if we use frequency analysis on instances of log events that belong to the same group. We use a simple pattern recognition technique based on text similarity to identify these groups. This contrasted with techniques that use clustering alone (e.g., AEL, and IPLOM) and those that apply frequency analysis on the entire log file (e.g., Logram). From this perspective, AWSOM-LP combines the best of these methods.

## 3 APPROACH

AWSOM-LP consists of three main steps: Pre-processing, pattern recognition, frequency analysis, and replacement of the remaining numerical dynamic variables. Similar to existing log parsing techniques, AWSOM-LP requires an initial dataset to recognize the structure of the log events. It goes through different steps to build a model that characterizes the information in the given log dataset. The first step is a pre-processing step where header information such as the timestamp, the log level, and the logging function are identified. We also replace trivial dynamic variables such as IP and MAC addresses by the expression '<*>'. The second

```
1   081109 203615 148 INFO dfs.DataNode$PacketResponder: PacketResponder 1 for block
    blk_38865049064139660 terminating
2   081109 203807 222 INFO dfs.DataNode$PacketResponder: PacketResponder 0 for block
    blk_-6952295868487656571 terminating
3   081109 204005 35 INFO dfs.FsNameSystem: BLOCK* NameSystem.addStoredBlock: blockMap
    updated: 10.251.73.220:50010 is added to blk_7128370237687728475 size 67108864
4   081109 204015 308 INFO dfs.DataNode$PacketResponder: PacketResponder 2 for block
    blk_8229193803249955061 terminating
5   081109 208106 329 INFO dfs.DataNode$PacketResponder: PacketResponder 2 for block
    blk_-6670958622368987959 terminating
6   081109 204132 26 INFO dfs.FsNameSystem: BLOCK* NameSystem.addStoredBlock: blockMap
    updated: 10,251.43.115:50010 is added to blk_3050920557425079149 size 67105864
7   081109 204328 34 INFO dfs.FsNameSystem: BLOCK* NameSystem.addStoredBlock: blockMap
    updated: 10,251.203.80:50010 is added to blk_7688946331004732825 size 67105864
8   081109 201453 24 INFO dfs.FsNameSystem: BLOCK* NameSystem.addStoredBlock: blockMap
    updated: 10.250.11.85:50010 is added to blk_2377150260125000006 size 67108064
9   081109 204525 512 IFO dfs.DatatlodefPacketResponder: PacketResponder 2 for block
    blk_572492839787299681 terminating
10  081109 201655 556 INFO dfs.DataNode$PacketResponder: Received block
    blk_3587505140051952248 of size 67IO8864 from /10.251.42.84
11  081109 204722 567 INFO dfs.DataNode$PacketResponder: Received block
    blk_5407003568334525940 of size 67108864 from /10.251.214.112
12  081109 204815 653 INFO dfs.DataNode$PacketResponder: Received block
    blk_9212264480425680329 of size 67108864 from /10.251.214.111
```

Fig. 2. HDFS log events used as a running example

step of AWSOM-LP is to identify similar log events and group them into patterns, used later to distinguish between the static and the dynamic tokens. The next step is to apply frequency analysis locally to instances of each group to determine the static and dynamic tokens. We conjecture that the frequency of static tokens is considerably higher than that of dynamic tokens when frequency analysis is used for each group of log events. The last step consists of fine-tuning the result to further improve the parsing accuracy and this is by replacing the remaining numerical dynamic variables. We explain each step in more detail in the following subsections. To illustrate our approach, we use the sample log events from the HDFS log dataset shown in Figure 2, which is one of the datasets used to evaluate AWSOM-LP. We added a line number to each log event to help explain the approach.

### 3.1 Pre-processing

We start pre-processing the log events by identifying header information, which usually contains the timestamp, process ID, log level, and the logging function. This information is easily identifiable using simple regular expressions as shown in similar studies (e.g., [27]). In addition, the LogPai datasets (used in the evaluation section) come with many regular expressions to detect headers in various log files. For example, in HDFS log events of Figure 2, we can see that all log events start with a timestamp (e.g., 081109 203615), a process ID (e.g., 148), a log level (e.g., INFO), and a logging function (e.g., dfs.DataNode$PacketResponder:). The regular expression that extracts this header information is as follows:

<Date> <Time> <Pid> <Level> <Component>: <Content>

Another essential part of the pre-processing step is the identification of trivial dynamic variables such as IP and MAC addresses and replace them with a standard token, which is <*> in our case. Identifying trivial variables can increase parsing accuracy as shown by He et al. [44] and Dai et al. [27]. This step also increases our chances of identifying similar log events that should be instances of the same pattern. The pattern recognition step is discussed in the next subsection.

In this paper, we detect the most common formats of the variables listed below (note that AWSOM-LP allows users to define other regular expressions that may describe, for example, domain-specific trivial variables, etc... ). The regular expressions to detect these variables are included in the AWSOM-LP git repository.[4]

- Directory paths such as "Library/Logs"
- IPv4 addresses with or without the port number such as "210.245.165.136" and "210.245.165.136:8080"
- Any value that starts with "0x" in the form of "0x0001FC".
- MAC addresses in the form of "FF:F2:9F:16:EB:27:00:0D:60:E9:14:D8"
- Months such as "Jan" or "January"
- Days such as "Thu" or "Thursday"
- Time such as "12:23:34.893"
- URL such as "http://www.google.com" and "https://www.google.com" (i.e., with https)

The result of applying the pre-processing step to the HDFS running example is shown in Figure 3 where the header information is omitted and the IP addresses in Lines 3, 6, 7, 10, 11, and 12 have been replaced by <*>.

4. https://github.com/SRT-Lab/awsom-lp

```
 1  PacketResponder 1 for block blk_38865049064139660 terminating
 2  PacketResponder 0 for block blk_-6952295868487656571 terminating
 3  BLOCK* NameSystem.addStoredBlock: blockMap updated: <*> is added to
    blk_7128370237687728475 size 67108864
 4  PacketResponder 2 for block blk_8229193803249955061 terminating
 5  PacketResponder 2 for block blk_-6670958622368987959 terminating
 6  BLOCK* NameSystem.addStoredBlock: blockMap updated: <*> is added to
    blk_3050920557425079149 size 67105864
 7  BLOCK* NameSystem.addStoredBlock: blockMap updated: <*> is added to
    blk_7688946331004732825 size 67105864
 8  BLOCK* NameSystem.addStoredBlock: blockMap updated: <*> is added to
    blk_2377150260125000006 size 67108064
 9  PacketResponder 2 for block blk_572492839787299681 terminating
10  Received block blk_3587505140051952248 of size 67I08864 from <*>
11  Received block blk_5407003568334525940 of size 67108864 from <*>
12  Received block blk_9212264480425680329 of size 67108864 from <*>
```

Fig. 3. Results of pre-processing HDFS log events example

### 3.2 Pattern Recognition

The second step of AWSOM-LP is to group similar log events into patterns. This grouping will help us later distinguish between the static and dynamic tokens by applying frequency analysis on the instances of each pattern. For example, the log messages of Lines 1, 2, 4, 5, and 9 all deal with terminating PacketResponder (used by HDFS to manage the processing of data into a series of pipeline nodes) and only vary in terms of the task number and the block id. So if we can group these log messages into the same pattern, we can easily see that the static tokens ("PacketResponder", "for", "block", and "terminating") appear more frequently than the dynamic tokens (0, 1, 2, and the specific block ids), hence the idea of using frequency analysis on instances of the same pattern instead of applying it to the entire log file, which may not lead to such a clear demarcation. Our grouping strategy relies on a simple string matching technique. More precisely, we measure the similarity of two log messages, $L_1$ and $L_2$, by counting the number of letters in $L_1$ and $L_2$ alphabetical terms (i.e., terms that do not contain any digits or special characters). The idea is to favor a comparison that makes most use of static tokens instead of dynamic tokens, which tend to vary more frequently across log events, misleading the matching process.

Assume $count(L_i)$ returns the number of letters in $L_i$ alphabetical terms, the similarity between two terms, $L_1$ and $L_2$ is measured as follows:

$$similarity(L_1, L_2) \& = \frac{count(L_1)}{count(L_2)}$$

For example, for $L_1$ = "PacketResponder <*> for block blk_38865049064139660 terminating" and $L_2$ = "PacketResponder <*> for block blk_-6952295868487656571 terminating", the similarity of $L_1$ and $L_2$ = 100%.

We need a threshold to assess the extent to which two log messages are deemed similar. In this paper, we use a 100% threshold to avoid any bias. This threshold is flexible and can be changed by the users.

In addition, AWSOM-LP is implemented in a way that any other similarity measure can be used. In our early experiments, we used cosine similarity by treating each log message as a vector of bag of words. However, we found that string matching offers a better compromise between accuracy and efficiency. In fact, one may also think of adopting a similarity measure that fits best a particular log file, without compromising efficiency. We defer this point to future work.

Applying the pattern recognition step to the log messages of Figure 3 results in three patterns. The first one consists of log messages 1, 2, 4, 5, and 9, which contain the static token PacketResponder. The second pattern consists of log messages 3, 6, 7, and 8, representing the message "BLOCK* NameSystem.addStoredBlock: blockMap update". The last one contains log messages 10, 11, and 12 for the "Received block" log event.

### 3.3 Frequency analysis

The next step is to apply frequency analysis to the instances of each pattern that were detected in the previous step to distinguish between static and dynamic tokens. To achieve this, we count the number of occurrences of each term in all instances of the same pattern. As discussed earlier, we conjecture that the static tokens appear more frequently than the dynamic ones within each pattern. For example, using the instances of the first pattern shown below we can see that the terms "PacketResponder", "for", "block", and "terminating" occur considerably more than individual block ids.

1) PacketResponder 1 for block blk_38865049064139660 terminating
2) PacketResponder 0 for block blk_-6952295868487656571 terminating
3) PacketResponder 2 for block blk_8229193803249955061 terminating
4) PacketResponder 2 for block blk_-6670958622368987959 terminating
5) PacketResponder 2 for block blk_572492839287299681 terminating

TABLE 1
Example of a frequency analysis results applied to Pattern 1

| Term | Frequency |
|---|---|
| PacketResponder | 5 |
| 0 | 1 |
| 1 | 1 |
| 2 | 3 |
| for | 5 |
| block | 5 |
| blk_38865049064139660 | 1 |
| blk_-6952295868487656571 | 1 |
| blk_8229193803249955061 | 1 |
| blk_-6670958622368987959 | 1 |
| blk_572492839287299681 | 1 |
| terminating | 5 |

We need a threshold in order to determine whether a term is considered static or not. For this, we use the minimum frequency. In other words, any token that appears strictly more than the minimum frequency is considered as a static token. We are aware that by doing so, we may misclassify some dynamic tokens as static. Another approach would be to take the median at the risk of losing some static tokens. It is not easy to find a threshold that yields the best trade-off. In this paper, we made the design choice to keep as many static tokens as possible, hence selecting the minimum frequency. In practice, this threshold can always be adjusted based on analyzing a sample log data. We can also resort to statistical methods to automatically determine the best threshold as proposed by Dai et al. [27], the creators of Logram, when applying frequency analysis to the entire log file. This said, one should be careful not to make the approach too complex, which may hinder its applicability. Our experiments with 16 log datasets (see the Evaluation section) show that the minimum frequency yields excellent results compared to all existing approaches including Logram.

Table 1 shows the frequency of the terms of the log messages of the first pattern from the running example. The minimum frequency is 1, which allows distinguishing between all static tokens ("PacketResponder", "block", "for", "terminating") from all the dynamic tokens except for dynamic token 2, which appears three times in the example, i.e. above the minimum frequency.

When we apply local frequency analysis to instances of all three patterns, the resulting log templates are shown below. We detected all static tokens and most dynamic tokens except 2 and 67108864.

1) PacketResponder <*> for block <*> terminating
2) BLOCK* NameSystem.addStoredBlock: blockMap updated: <*> is added to <*> size 67108864
3) PacketResponder 2 for block <*> terminating
4) Received block <*> of size 67108864 from <*>

### 3.4 Replacing Remaining Numerical Variables

The last step of AWSOM-LP is to fine-tune the algorithm by replacing the remaining numerical dynamic variables that were not identified during the previous step. More precisely, we automatically consider any numbers that appear between spaces, parentheses, or brackets (e.g., " 980 ", "(980)" and "[678]" as dynamic tokens. One might think that this step could have been included as part of the pre-processing stage of AWSOM-LP when looking for trivial variables. However, we found that doing so will affect the result of the frequency analysis step by letting more non-numerical dynamic variables appear more than the threshold, ending up falsely included as static tokens.

The final result of parsing the HDFS example log events of Figure 2 is shown below. As we can see all templates have accurately been detected.

1) PacketResponder <*> for block <*> terminating
2) BLOCK* NameSystem.addStoredBlock: blockMap updated: <*> is added to <*> size <*>
3) Received block <*> of size <*> from <*>

## 4 EVALUATION

In this section, we evaluate the effectiveness of AWSOM-LP in parsing logs of 16 log datasets of the LogPai benchmark [30] available online[5]. The logs were generated from various systems including Apache, Android, HDFS, Linux, and so on, as shown in Table 2. Other studies have used the same LogPai benchmark. We assessed AWSOM-LP's accuracy based on manually labelled log data files where log events are mapped out to log templates. This labelling was provided by the LogPai team as part of the LogPai project.

TABLE 2
LogPai datasets

| Datasets | Description | Size |
|---|---|---|
| Apache | Apache server error log | 5.1MB |
| Android | Android framework log | 192MB |
| BGL | Blue Gene/L supercomputer log | 743MB |
| HDFS | Hadoop distributed file system log | 1.47GB |
| Hadoop | Hadoop mapreduce job log | 2MB |
| HPC | High performance cluster | 32MB |
| Linux | Linux system log | 2.25MB |
| Mac | Mac OS log | 17MB |
| OpenSSH | OpenSSH server log | 73MB |
| Proxifier | Proxifier software log | 2.42MB |
| Spark | Spark job log | 166MB |
| HealthApp | HealthApp log | 24MB |
| Thunderbird | Thunderbird supercomputer log | 29.60GB |
| Windows | Windows event log | 26.09GB |
| OpenStack | OpenStack software log | 41MB |
| Proxifier | Proxifier software log | 3MB |
| Zookeeper | ZooKeeper service log | 10MB |

We followed the evaluation protocol provided by Zhu et al. [30], which focuses on three aspects: Accuracy, efficiency, and ease of stabilisation.

- Accuracy: The accuracy of a log parser is defined as its ability to recognize static tokens and dynamic tokens in each log event, and associate log events to the correct log template. He et al. [44] showed the necessity of high log parsing accuracy, especially avoiding false positives, to increase the confidence and reliability in using a given log parser.

5. https://zenodo.org/record/3227177#.YUqmXtNPFRE

- Efficiency: Because of the considerable number of logs that are generated, a log parser must be efficient by having a reasonable running time. This is particularly important since log parsing is the first phase of log analysis. Having a slow log parser may create delays in uncovering key knowledge from logs, which in turn may deter users from using the tool.
- Ease of stabilisation: In order to distinguish between the static and dynamic components in a log message, most existing log parsers, including AWSOM-LP, use an initial dataset to recognise the structure of the logs. It is desirable for a log parser to obtain stable results after acquiring knowledge from a limited number of existing logs, so that parsing logs may be done in real time without the need to update this knowledge.

### 4.1 Accuracy

Each log dataset of the LogPai benchmark used in this study comes with a subset of 2,000 log events that have been parsed manually by the LogPai team. For a given log dataset, a number of log templates have been identified and each log event out of the 2,000 events has been associated to a specific log template. This ground truth is meant for researchers to test their log parsers and has been used by many log parsing tools such as Drain [29], AEL [31], Lenma [28], IPLoM [33], and the latest approach, Logram [27]. We also use it in this study to evaluate AWSOM-LP and to compare AWSOM-LP with existing tools. Table 3 shows an example of events from the Apache log dataset where a log event (represented here by an id starting with "E") is mapped to a template.

To measure accuracy, we follow the same metrics as the ones used in related studies (e.g., Logram), which assess the ability of their tool to group together similar log messages that were identified as part of the same template in the ground truth. For this reason, they measure accuracy by examining pairs of log messages to see whether they can be detected as part of the same template according to the ground truth. More precisely, they use precision, recall, and F1-score (which is the accuracy), defined as follows:

$$Precision = \frac{TP}{TP + FP}$$

$$Recall = \frac{TP}{TP + FN}$$

$$Grouping\,Accuracy(F1\_Score) = \frac{2 \times Precision \times Recall}{Precision + Recall}$$

- True positive (TP): The number of pairs of log messages that are correctly parsed as instances of the same template according to the ground truth.
- False positive (FP): The number of pairs of log messages that are incorrectly parsed as instances of the same template according to the ground truth.
- False negative (FN): The number of pairs of log messages that should be parsed as instances of the same template, but each belongs to a different template.

TABLE 3
An example of manually labeled log events from the Apache log dataset

| Event ID | Event Template |
|---|---|
| E1 | jk2_init() Found child <*>in scoreboard slot <*> |
| E2 | workerEnv.init() ok <*> |
| E3 | mod_jk child workerEnv in error state <*> |
| E4 | [client <*>] Directory index forbidden by rule: <*> |
| E5 | jk2_init() Can't find child <*>in scoreboard |
| E6 | mod_jk child init <*><*> |

However, this evaluation method only assesses whether log events of similar structure were identified as part of the same template according to the ground truth. It does not guarantee that the static and dynamic tokens of log events were properly recognized. By analyzing the parsing output of some log parsers (e.g., AEL and Logram), we found many cases where the static and dynamic variables contained in the log messages were not correctly recognized and yet this was not reflected in the parsing accuracy. For example, the log event:

job_1445144423722_0020Job Transitioned from NEW to INITED

from Hadoop benchmark was parsed by Logram into :

<*> Transitioned from NEW to <*>

We can see that the string "INITED" was mistakenly detected as a dynamic variable based on the ground truth, where it is considered as a static token. This is because job transitions in Hadoop contain a finite number of states, which provide developers critical information on how to debug problems related to job scheduling. The ground truth log parsing file lists the possible Hadoop job transitions as three distinct templates, which are:

job<*>Job Transitioned from INITED to SETUP
job<*>Job Transitioned from NEW to INITED
job<*>Job Transitioned from SETUP to RUNNING

In this paper, in addition to grouping accuracy, we introduce an additional metric that we call "matching accuracy", in which we measure the ratio of the number of log events for which we correctly identified the static and dynamic tokens to the number of log events parsed. We believe that this metric reflects better the ability of a log parser to recognize log events.

To calculate the matching accuracy metric, we implemented a script to verify for each log message's parsing result to see if the static and dynamic tokens are accurately processed. The script checks that the parsed line and the manually labelled log line from the dataset contain:

- The same static tokens (we also make sure the tokens appear in the right order), and
- The dynamic variables are correctly identified.

*Results*

Table 4 shows the results of AWSOM-LP grouping accuracy and a comparison with other tools. **AWSOM-LP has the best accuracy in parsing 13 of the 16 log datasets**, in comparison to other log parsing tools including the recent one, Logram. Additionally, our approach has an **average accuracy of**

TABLE 4
Accuracy of AWSOM-LP compared with other log parsers. We have highlighted the results that are the highest among the parsers in bold.

| Name | Drain | AEL | Lenma | IPLoM | Logram | AWSOM-LP |
|---|---|---|---|---|---|---|
| Android | 0.933 | 0.867 | 0.976 | 0.716 | 0.848 | **0.970** |
| Apache | 0.693 | 0.693 | 0.693 | 0.693 | 0.699 | **0.999** |
| BGL | 0.822 | 0.818 | 0.577 | 0.792 | 0.740 | **0.945** |
| Hadoop | 0.545 | 0.539 | 0.535 | 0.373 | 0.965 | **0.991** |
| HDFS | 0.999 | 0.999 | 0.998 | 0.998 | 0.981 | **0.988** |
| HealthApp | 0.609 | 0.615 | 0.141 | 0.651 | **0.969** | 0.955 |
| HPC | 0.929 | 0.990 | 0.915 | 0.979 | 0.959 | **0.997** |
| Linux | 0.250 | 0.241 | 0.251 | 0.235 | 0.460 | **0.988** |
| Mac | 0.515 | 0.579 | 0.551 | 0.503 | 0.666 | **0.977** |
| openSSH | 0.507 | 0.247 | 0.522 | 0.508 | 0.847 | **0.945** |
| Openstack | 0.538 | 0.718 | 0.759 | 0.697 | 0.545 | **0.840** |
| Proxifier | 0.973 | 0.968 | 0.955 | **0.975** | 0.951 | 0.739 |
| Spark | 0.902 | 0.965 | 0.943 | 0.883 | 0.903 | **0.992** |
| Thunder... | 0.803 | 0.782 | **0.814** | 0.505 | 0.761 | 0.669 |
| Windows | 0.983 | 0.983 | 0.277 | 0.554 | 0.957 | **0.984** |
| Zookeeper | 0.962 | 0.922 | 0.842 | 0.967 | 0.955 | **0.999** |
| Average | 0.748 | 0.745 | 0.672 | 0.689 | 0.825 | **0.936** |

TABLE 5
Comparison between AWSOM-LP grouping and matching accuracy

| Name | Grouping Accuracy | Matching Accuracy |
|---|---|---|
| Android | 0.970 | 0.879 |
| Apache | 0.999 | 0.990 |
| BGL | 0.945 | 0.437 |
| Hadoop | 0.991 | 0.427 |
| HDFS | 0.988 | 0.997 |
| HealthApp | 0.955 | 0.887 |
| HPC | 0.985 | 0.997 |
| Linux | 0.988 | 0.839 |
| Mac | 0.977 | 0.652 |
| openSSH | 0.945 | 0.801 |
| Openstack | 0.840 | 0.816 |
| Proxifier | 0.739 | 0.810 |
| Spark | 0.992 | 0.755 |
| Thunderbird | 0.669 | 0.363 |
| Windows | 0.984 | 0.694 |
| Zookeeper | 0.999 | 0.683 |
| Average | 0.936 | 0.730 |

**0.936**, whereas the second most accurate method, Logram, has an average accuracy of 0.825. Further, **AWSOM-LP has an accuracy of more than 0.9 in 13 out of 16 log datasets**, demonstrating the excellent ability of AWSOM-LP to parse various log files generated from different systems. We carefully examined the cases that AWSOM-LP missed and found that the causes are as follows:

- We found errors and inconsistencies in the manual labelling of the benchmark files. For example, in the Mac log file, the token 'CrazyIvan46!' in log event 'labeling -[NETClientConnection evaluateCrazyIvan46] CI46 - Perform CrazyIvan46!' is considered as static, which is an error in the labelling of the data. This is a dynamic token, which refers to a username. An example of an inconsistency would be the case of the HDFS log dataset where the block ids such as blk_23333989 sometimes labelled as <*> and some other times as blk_<*>.
- Dynamic variables that appear more than the minimum frequency, and therefore are considered as static tokens. For example, the variable '2607:f140:6000:8:c6b3:1ff:fecd:467f' appears 18 times in the Mac log file and six times locally (in a group of 20 similar log events). Even by using local frequency, its number of occurrences remains higher than the threshold.
- There are many static variables, which are syntactically similar to dynamic variables. For example, in the Hadoop log file, the log event '2015-10-18 18:01:51,963 INFO [main] org.mortbay.log: jetty-6.1.26' is interpreted by AWSOM-LP as '<*>' after pre-processing the other fields and the pattern recognition step, which is incorrect according to the manually labeled files. The term jetty-6.1.26 should be a static token. This type of static tokens are the hardest to detect because they bear most of the characteristics of dynamic variables.
- We found dynamic variables that can be interpreted in different ways. Take, for example, the IP address 10.251.73.220:50010. It can be interpreted either as one dynamic variable or two variables consisting of the the IP address '10.25.73.220' and the port number '50010'.

The matching accuracy of AWSOM-LP is shown in Table 5. Unfortunately, we could not compare matching accuracy to the other tools since previous studies did not calculate this metric. AWSOM-LP's matching accuracy for Apache and HDFS logs is more than 99%, which means that most of the static and dynamic tokes of their log events have been correctly identified and each log event was mapped out to the proper log template. This metric is more than 80% in 9 out of the 16 log datasets. The average matching accuracy is 0.73, making AWSOM-LP a very reliable parsing tool.

One of the main reasons for the low matching accuracy for BGL, Hadoop and Thunderbird datasets is due to the errors in the manual labelling of the ground truth file. For example, the log event "10722 total interrupts. 0 critical input interrupts. 0 microseconds total spent on critical input interrupts, 0 microseconds max time in a critical input interrupt" is manually labelled as " <*> total interrupts. 0 critical input interrupts. <*> microseconds total spent on critical input interrupts, <*> microseconds max time in a critical input interrupt." AWSOM-LP generates the following pattern " <*> total interrupts. <*> critical input interrupts. <*> microseconds total spent on critical input interrupts, <*> microseconds max time in a critical input interrupt." Then only difference is in the manually labelled pattern we have " 0 critical input interrupts", where the number 0 is considered as a static token in the ground truth, whereas AWSOM-LP deemed it as a dynamic variable. This mismatch appears 71 times in the log dataset. Another example is the log event "generating core.21045" which is manually labelled as " generating core<*> ", and detected by AWSOM-LP associate that event with the template "generating <*>. The fact that the variable "core.21045" does not contain a white space that separates " core" from "21045" made AWSOM-LP interpret it as one dynamic variable. This mismatch appears 721 times in the log dataset. Other types of mistakes between the result of AWSOM-LP and the ground trught are shown in Table 6.

TABLE 6
Examples of mismatches between the ground truth and AWSOM-LP

| Log file | Comparison |
|---|---|
| BGL | Event: generating core.21045<br>Ground truth : generating core<*><br>AWSOM-LP : generating <*> |
| BGL | Event: 10765 total interrupts. 0 critical input interrupts.<br>Ground: <*>total interrupts. 0 critical input interrupts.<br>AWSOM-LP : <*>total interrupts. <*>critical input interrupts. |
| BGL | Event : idoproxydb hit ASSERT condition:<br>ASSERT expression=0<br>Ground truth : idoproxydb hit ASSERT condition:<br>ASSERT expression=0<br>AWSOM-LP : idoproxydb hit ASSERT condition:<br>ASSERT expression=<*> |
| Hadoop | Event : task_1445144423722_0020_m_000007 Task Transitioned from NEW to SCHED<br>Ground: task<*>Task Transitioned from NEW to SCHED<br>AWSOM-LP : <*>Task Transitioned from NEW to SCHED |
| Hadoop | Event : Progress of TaskAttempt attempt_1445144423_0020_m_000000_0 is : 0.1795899<br>Ground: Progress of TaskAttempt attempt<*>is :<*><br>AWSOM-LP : Progress of TaskAttempt <*>is :<*> |
| Hadoop | Event. : Using callQueue class java.util.concurrent.LinkedBlockingQueue<br>Ground: Using callQueue class java.util.concurrent.LinkedBlockingQueue<br>AWSOM-LP : Using callQueue class <*> |
| Thunder | Event : (Release Date: Mon Sep 27 22:15:07 EDT 2004)<br>Ground truth : <*>(Release Date: <*>EDT <*>)<br>AWSOM-LP : <*>(Release Date: <*>) |

### 4.2 Ease of stabilisation

We assess the ease of stabilization by processing portions of the logs and measure how much knowledge in terms of input log events is required by AWSOM-LP to reach a reasonable accuracy. In other words, we want to know the likelihood of creating a comprehensive set of log templates from a small amount of log data that can recognize future raw log events. Building a dictionary of templates from a limited set of log events is desirable for improved efficiency and scalability. To measure ease of stabilization, we follow the same approach as the one proposed by Dai et al. [27]. For each dataset, we select portions of the log file and measure how many templates are extracted using these portions compared to the total number of templates that are generated when using the entire file. We start with 5% of log events and increase this portion by 10% until we reach 100% coverage of the log file. This part of the evaluation is applied to larger log files (not only the 2,000 log events as for accuracy). The LogPai log datasets are organized in two different ways. Some datasets such as Hadoop and OpenStack are organized into folders that contain multiple log files. For these, we selected the largest log file from each dataset for evaluation. The other datasets such as BGL, HDFS, etc. are saved as one log file. For these, we took the entire file as our testbed. However, we excluded Thunderbird, Windows, and HDFS because of the large amount of data they contain, which is in Giga Bytes. Additionally, we excluded the HPC dataset because the file was corrupt and we were not able to process it. The exact files used to evaluate ease of stabilization can be found on the AWSOM-LP Github repository. In total, we used 12 log files out of the 16 datasets of LogPai.

*Results*

Figure 4 shows the results of ease of stabilization of AWSOM-LP. The red line in each figure means that 80% of the templates have been detected. As we can see, AWSOM-LP can generate 80% of log templates with less than 35% of the logs in 8 log datasets out of 12. The results are even better with Apache, Mac, Spark, and Proxifier log files in which case AWSOM-LP needs less than 10% of the logs to generate 80% of the templates. For Hadoop and HealthApp logs, we need 25% of the total size of the log files. For BGL, OpenSSH, Android, and Openstack, AWSOM-LP requires at least 50% of the data to discover 80% and more templates. These results can be attributed to the fact that these logs contains a large number of patterns, forcing AWSOM-LP to learn new patterns as new logs are coming in. We did not find a correlation between ease stabilisation and the size of log files. Smaller log files such as Hadoop (7,000 log events) may require a large set to stabilize AWSOM-LP than larger files. The key factors that impact ease of stabilization are the number of patterns they contain and the high variability in the data (i.e., newer log events are introduced frequently during the execution of the logged system).

### 4.3 Efficiency

To assess the efficiency of AWSOM-LP, we record the execution time to complete the end-to-end parsing process. For this aspect of the study, we use the same log files as the ones used to assess ease of stabilization (see Table 7). Note that for Android, BGL, and Spark logs, which contains more than 1 million events, we measure AWSOM-LP's efficienct for these logs up to 1 million log events. This limitation is mainly caused by our limited computing environment. All experiments were conducted on a MacBook Pro Laptop running MacOS Big Sur version 11.4 with a 6 Intel Core i7 CPU 2.2GHz, 32GB 2400MHz DDR4 RAM, and 256 GB SSD hard drive.

Unlike other studies that use the file size to measure efficiency [30] [45], we decided in this paper to use the number of log events instead. We believe that the number of log events is more representative of the amount of information contained in a log file. For example, the Hadoop log file has a size of 1.5MB, but only contains 7,253 log events, whereas the size of the Spark log file is 1MB, but has 8,750 log events. This because the format and complexity of log events vary from one dataset to another. Using the size of the storage space to measure efficiency is misleading.

TABLE 7
The log datasets used for ease of stabilisation

| File | Size | Number of Lines |
|---|---|---|
| Apache | 5.1MB | 56,482 |
| Android | 192.3MB | 1,555,000 |
| BGL | 743.2MB | 4,747,000 |
| Spark | 166MB | 1,225,000 |
| OpenSSH | 73MB | 655,000 |
| Zookeeper | 10MB | 74,000 |
| OpenStack | 41MB | 137,000 |
| Proxifier | 3MB | 21,000 |
| Mac | 17MB | 117,000 |
| HealthApp | 24MB | 253,000 |
| Hadoop | 2MB | 7,000 |
| Linux | 2.4MB | 27,000 |

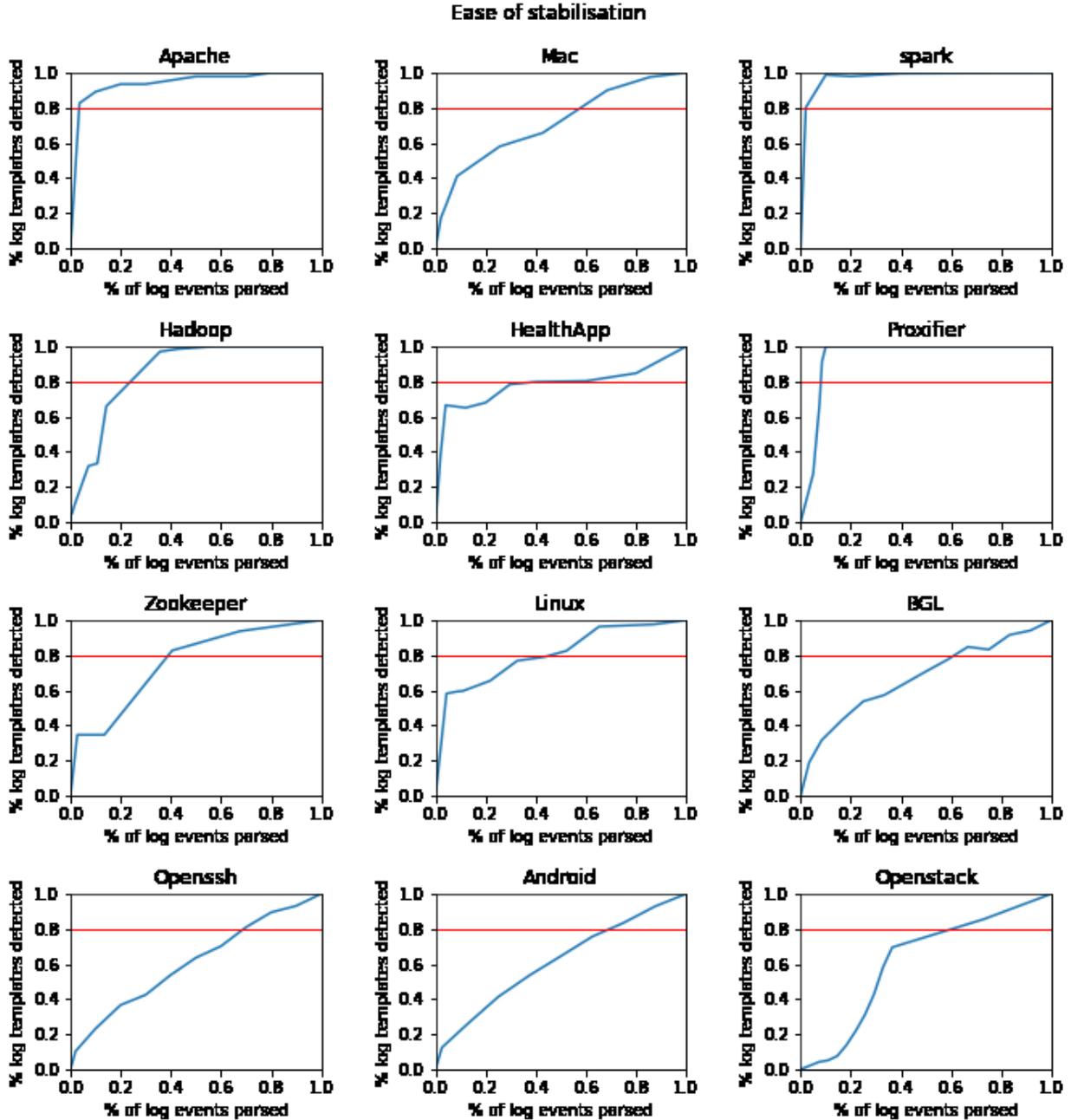

Fig. 4. Results of ease of stabilization of AWSOM-LP. The x-axis refers to the percentage of log events used for parsing, while the y-axis refers to the percentage of templates that were identified. The red line indicates that 80% of the templates were identified.

For each file, we measure the efficiency of AWSOM-LP by randomly selecting various portions of the file to see how AWSOM-LP performs as the size of the file increases. For example, for the Apache log file of Table 7, we start by measuring the efficiency of LogPaser when applied to 10,000 log events, and increase this number by a factor of 2 until we reach the total size of the file (i.e., 56,482).

*Results*

Figure 5 shows the results of AWSOM-LP's efficiency, which varies from 3 seconds to 1,296 seconds (i.e., 22 min) depending on the log file. For Hadoop, Proxifier, Apache, Linux, Zookeper, MAC, AWSOM-LP took from 3 to 100 seconds to parse each file. For larger logs such as HealthApp, Spark, OpenSSH, BGL, Android, AWSOM-LP took between

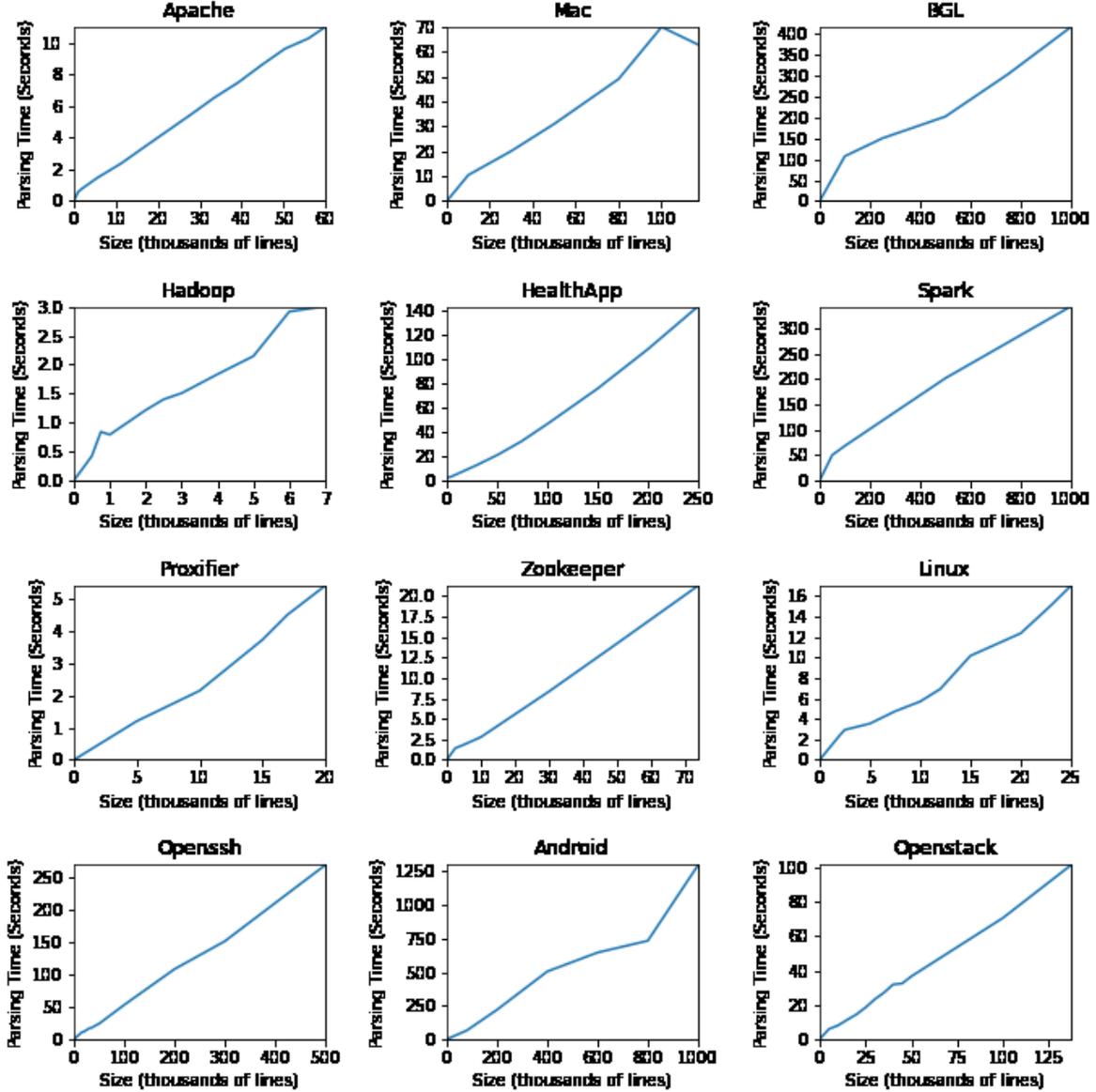

Fig. 5. Results of AWSOM-LP efficiency. The x-axis represents the number of log events and the y-axis represents the execution time in seconds.

143 (3 min) to 1,296 seconds (22 min). This variation is mainly due to the structure of the log events of each file and of course the number of lines processed. Note, however, that Android logs took the most time to parse 1,296 seconds (22 min) and this is due to the complexity of Android events. Each event contains a large number of static and dynamic tokens, which affect the AWSOM-LP's pattern recognition step. In average, AWSOM-LP's efficiency is 233 seconds (4 min).

Additionally, the figure shows that the parsing time increases linearly with respect to the file size (in terms of the number of log events), meaning that AWSOM-LP's efficiency does not degrade as the size of the file increases.

In conclusion, the overall efficiency of AWSOM-LP is very good, which is 5 minutes on average. This performance is obviously impacted by the hardware used to run the experiments. We anticipate that using more computing power and better hardware, combined with parallel programming, would significantly increase the efficiency of AWSOM-LP.

## 5 THREATS TO VALIDITY

In this section, we discuss the threats to the validity of this study, which are organized as internal, external, conclusion, and reliability threats.

**Internal validity:** Internal validity threats concern the factors that might impact our results. We assessed AWSOM-LP using 16 log datasets from the LogPai benchmark.

We cannot ascertain that AWSOM-LP's accuracy would be the same if applied to other datasets. This said, these datasets cover software systems from different domains, which make them a good testbed for log parsing and analysis tools. Another internal threat to validity is related to the threshold we used when applying local frequency analysis, which is the minimum frequency. A different threshold may lead to different results. To mitigate this threat, we experimented with different threshold including the median frequency and found that the minimum frequency yields best results. We should work towards automating the selection of the threshold tailored to specific log files based on statistical analysis of sample log data. Also, despite our efforts implementing and testing AWSOM-LP, errors may have occurred. To mitigate this threat, we tested the tool on many log files and we also checked manually its output on small samples. In addition, we make the tool and the data available on Github to allow researcher to reproduce the results. Finally, to check the accuracy of AWSOM-LP, we have had to examine the differences between the results obtained by AWSOM-LP and the ground truth. This was done semi-automatically through scripts and manual inspections. All efforts were made to reduce potential errors.

**Reliability Validity**. Reliability validity concerns the possibility to replicate this study. We provide an online package to facilitate the assessment, replicability and reproducibility of this study.

**Conclusion validity.** Conclusion validity threats correspond to the correctness of the obtained results. We applied AWSOM-LP to 16 log files that are widely used in similar studies. We made every effort to review the accuracy (grouping and matching), efficiency, and ease of stabilisation experiments to ensure that we properly interpret the results. The tool and the files used in every step of this study are made available online to allow the assessment and reproducibility of our results.

**External validity:**. External validity is about the generalizability of the results. We performed our study on 16 log files that cover a wide range of software systems. We do not claim that our results can be generalized to all possible log files, in particular to industrial and proprietary logs to which we did not have access.

## 6 CONCLUSION

We presented AWSOM-LP, a powerful log parsing approach and tool that can be used by researchers and practitioners to parse and abstract unstructured raw log data, an important first step of any viable log analysis task. AWSOM-LP differs from other tools in its design. It uses a clever way to distinguish between static and dynamic tokens of log events by applying frequency analysis to instances of events that are grouped in the same pattern. By doing so, AWSOM-LP is capable to clearly extract log templates that can be used to recognized and structure log events. AWSOM-LP is more accurate in parsing a representative set of 16 log files of the LogPai project than any existing open source log file. Not only that, AWSOM-LP is also efficient. It took in average 4 min to parse 12 log file with up to 1 million events. Further, AWSOM-LP does not need to read the entire log file to learn the templates. IT requires between 10% to 50% of the data to recognize at least 80% of the template, making it a very stable tool, ready to process logs in real-time.

Future work should build on this work by focusing on the following aspects (a) apply AWSOM-LP to more logs, especially those from industrial system, (b) improve AWSOM-LP by adding more regular expressions to identify other trivial dynamic variables such as domain-specific variables, (c) investigate a simple way to recommend a cutoff threshold for the frequency analysis step based on the characteristics of the log data, and (d) improve the efficiency of the tool when applied to log files with a large number of patterns, with high variability.